# Counterexample-Guided Polynomial Loop Invariant Generation by Lagrange Interpolation

Yu-Fang Chen[1], Chih-Duo Hong[1], Bow-Yaw Wang[1], and Lijun Zhang[2]

[1] Institute of Information Science, Academia Sinica, Taiwan
[2] State Key Laboratory of Computer Science, Institute of Software, CAS, China

**Abstract.** We apply multivariate Lagrange interpolation to synthesizing polynomial quantitative loop invariants for probabilistic programs. We reduce the computation of a quantitative loop invariant to solving constraints over program variables and unknown coefficients. Lagrange interpolation allows us to find constraints with less unknown coefficients. Counterexample-guided refinement furthermore generates linear constraints that pinpoint the desired quantitative invariants. We evaluate our technique by several case studies with polynomial quantitative loop invariants in the experiments.

## 1 Introduction

A probabilistic program may change its computation due to probabilistic choices. Consider, for instance, the Miller-Rabin algorithm for primality test [27]. Given a composite number, the algorithm reports incorrectly with probability at most 0.25. Since the outcome of the algorithm is not always correct, classical program correctness specifications [9, 14, 20] do not apply. For probabilistic programs, quantitative specifications are needed to reason about program correctness [8, 23, 24]. Instead of logic formulae, probabilistic programs are specified by numerical functions over program variables. Since a probabilistic program gives random outcomes, a numerical function may have different values on different executions. The expected value of a numerical function is then determined by the probability distribution induced by the executions of program.

Since probabilistic programs are specified by numerical functions, their correctness can be established by annotations with expectations. In particular, correctness of while loops can be proved by inferring special expectations called the *quantitative loop invariants* [24, 25]. Similar to classical programs, finding general quantitative loop invariants is hard. Techniques for generating linear quantitative loop invariants however are available [1, 15, 22, 25].

Interestingly, existing linear loop invariant generation techniques can be extended to synthesize polynomial invariants [1]. Observe that polynomial multivariate polynomials are linear combinations of monomials. For instance, any polynomial over $x, y$ with degree 2 is a linear combination of the monomials $1, x, y, x^2, y^2$, and $xy$. It suffices to find coefficients of the monomials to represent any multivariate polynomial of a fixed degree. Linear loop invariant generation techniques can hence be applied to infer invariants of a fixed degree. The number of monomials however grows rapidly. Quadratic polynomials over 5 variables, for example, are linear combinations of 21 monomials. One then has to find as many coefficients. It is unclear whether the extended approach is still feasible.

In this paper, we develop a Lagrange interpolation-based technique to synthesize polynomial loop invariants for simple loops in probabilistic programs. Lagrange interpolation is a well-known method to construct explicit expressions for polynomials by sampling. For example, suppose that the values of $f(x)$ are known to be $s_1$, $s_3$, and $s_4$ at the sampling points 1, 3, and 4, respectively. By Lagrange interpolation, we immediately have an explicit expression of $f(x) = s_1 \cdot \frac{(x-3)(x-4)}{(1-3)(1-4)} + s_3 \cdot \frac{(x-1)(x-4)}{(3-1)(3-4)} + s_4 \cdot \frac{(x-1)(x-3)}{(4-1)(4-3)}$. Our new technique employs multivariate Lagrange interpolation. Similar to previous techniques [22, 15], we use conditions of quantitative loop invariants as constraints. Lagrange interpolation moreover allows us to simplify the constraints and sometimes to determine several coefficients. In the example, suppose $f(3) = 1$ is known. Then it suffices to determine $s_1$ and $s_4$ to construct an explicit expression of $f(x)$. In contrast, if $f(x)$ is represented as $c_0 + c_1 x + c_2 x^2$, then $f(3) = 1$ only gives $c_0 + 3c_1 + 9c_2 = 1$ and determines none of the coefficients. Lagrange interpolation hence can reduce the number of unknown coefficients and make our technique more scalable.

Although there are less unknown coefficients, one still has to solve non-linear constraints. We give heuristics to determine coefficients efficiently. Our heuristics first perform random experiments and obtain linear constraints about coefficients. An SMT solver is then used to find candidate coefficients from the constraints. If there is no candidate, then the desired loop invariant does not exist. Otherwise, quantifier elimination verifies whether the candidate coefficients give a loop invariant. If so, our technique has found a quantitative loop invariant. Otherwise, we add more linear constraints to exclude infeasible coefficients.

We apply our technique to find quantitative loop invariants for ten annotated loops from non-trivial probabilistic programs. Our case studies range from gambler's ruin problem [13] to simulation of a fair coin with a biased coin [15]. Over 1000 random runs, our technique is able to synthesize polynomial quantitative loop invariants within 15 seconds on average. Besides, 97.5% of the runs can finish within a 300-second timeout.

*Related Work.* Constraint-based techniques for automated loop invariants generation have been much progressed over the past years [4, 5, 17, 18, 21, 22, 29]. Gupta and Rybalchenko [18, 19] proposed a CEGAR framework, so that static and dynamic information of a program can be exploited incrementally to restrict the search space of *qualitative* loop invariants. Sankaranarayanan *et al.* [29] used Gröbner bases to reduce the generation of algebraic polynomial loop invariants to solving non-linear constraints in the parametric linear form. These techniques however deal with classical programs and cannot be applied to probabilistic programs directly. McIver and Morgan [24] were among the first to consider *quantitative* loop invariants for probabilistic programs. Katoen *et al.* [22] studied the synthesis of quantitative loop invariants using a constraint-solving approach.

The approach was further developed and implemented in the PRINSYS tool [15], which synthesizes quantitative invariants by solving constraints over unknown template coefficients. The performance of the tool however is sensitive to manually supplied templates. Recently, a technique based on abstract interpretation is proposed in [1]. It formulates linear loop invariants with the collecting semantics and synthesizes coefficients via fixed-point computation. Although the authors only report experiments on linear loop invariants, the technique can be extended to generate polynomial invariants by representing polynomials as linear combinations of monomials. The effectiveness of the extension however is unclear.

We have the following organization. After preliminaries, we review probabilistic programs in Section 3. Quantitative loop invariants are presented in Section 4. Section 5 introduces multivariate Lagrange interpolation. Our technical contribution is presented in Section 6. Applications are given in Section 7. We evaluate our technique in the following section. Section 9 concludes our presentation.

## 2 Preliminaries

Let $\mathbf{x}_m$ be a sequence of variables $x_1, x_2, \ldots, x_m$. We use $\mathbb{R}[\mathbf{x}_m^n]$ to denote the set of real coefficient polynomials over $m$ variables of degree at most $n$. Observe that $\mathbb{R}[\mathbf{x}_m^n]$ can be seen as a vector space over $\mathbb{R}$ of dimension $d = \binom{m+n}{n}$. For instance, the set of $d$ monomials $\{x_1^{d_1} x_2^{d_2} \cdots x_m^{d_m} : 0 \leq d_1 + d_2 + \cdots + d_m \leq n\}$ forms a basis of $\mathbb{R}[\mathbf{x}_m^n]$. Given $f \in \mathbb{R}[\mathbf{x}_m^n]$ and expressions $e_1, e_2, \ldots, e_m$, we use $f(e_1, e_2, \ldots, e_m)$ to denote the polynomial obtained by replacing $x_i$ with $e_i$ for $1 \leq i \leq m$ in $f$. Particularly, $f(\mathbf{v})$ is the value of $f$ at $\mathbf{v} \in \mathbb{R}^m$.

A *constraint* is a quantifier-free logic formula with equality, linear order, addition, multiplication, division, and integer constants. A constraint is *linear* if it contains only linear expressions; otherwise, it is *non-linear*. A *quantified constraint* is a constraint with quantifiers over its variables. A *valuation* over $\mathbf{x}_m$ assigns a value to each variable in $\mathbf{x}_m$. A *model* of a constraint is a valuation which evaluates the constraint to true.

Given a quantified constraint, *quantifier elimination* removes quantifiers and returns a logically equivalent constraint. Given a linear constraint, a *Satisfiability Modulo Theory (SMT) solver* returns a model of the constraint if it exists.

## 3 Probabilistic Programs

A *probabilistic program* in the *probabilistic guarded command language* is of the following form:

$$P ::= \mathsf{skip} \mid \mathsf{abort} \mid x := E \mid P;P \mid P[p]P \mid \mathsf{if}\ (G)\ \mathsf{then}\ \{P\}\ \mathsf{else}\ \{P\} \mid \mathsf{while}\ (G)\ \{P\}$$

where $E$ is an expression and $G$ is a Boolean expression. For $p \in (0, 1)$, the *probabilistic choice command* $P_0[p]P_1$ executes $P_0$ with probability $p$ and $P_1$ with probability $1 - p$. For instance, $x := 1\,[0.75]\,x := 0$ sets $x$ to 1 with probability 0.75 and to 0 with probability 0.25. A *program state* is a valuation over program variables. For simplicity, we assume program variables are in non-negative integers, and use 0 and 1 for the truth values false and true respectively.

*Example 1.* Consider the following probabilistic program:

$$z := 0;\ \textsf{while}\,(0 < x < y)\,\{\ x := x + 1\,[0.5]\,x := x - 1;\ z := z + 1;\ \}$$

The program models a game where a player possesses $x$ dollars and keeps tossing a fair coin. The player wins one dollar for each head and loses one dollar for each tail. The game ends either when the player loses all his capital, or when the amount of capital reaches $y$ dollars for a predetermined $y \geq x$. The variable $z$ counts the number of tosses made during the game.

### 3.1 Expectations

From an initial program state, a probabilistic program can have different final program states due to probabilistic choice commands. Particularly, a function over program variables gives different values on different final program states. Note that a probabilistic program induces a probability distribution on final program states. One therefore can discuss the expected value of any function over program variables with respect to that probability distribution. More precisely, one can take an expectation transformer [24] approach to characterize a probabilistic program by annotating the program with expectations.

Formally, an *expectation* is a function mapping program states to a non-negative real number. An expectation is called a *post-expectation* when it is to be evaluated on final program states. Similarly, an expectation is called a *pre-expectation* if it is to be evaluated on initial program states. Let $preE$ and $postE$ be expectations, and $prog$ a probabilistic program. We say a *quantitative* Hoare triple $\langle preE \rangle\ prog\ \langle postE \rangle$ holds if the expected value of $postE$ is no less than that of $preE$ before executing $prog$. Note that the expected values of $postE$ and $preE$ are functions over states and hence are compared pointwisely.

For any Boolean expression $G$, define the *indicator* function $[G] = 1$ if $G$ is true and $[G] = 0$ otherwise. Consider an *qualitative* Hoare triple $\{P\}\ prog\ \{Q\}$ with a pre-condition $P$, a post-condition $Q$, and a classical program $prog$. Observe that $\{P\}\ prog\ \{Q\}$ holds if and only if $\langle [P] \rangle\ prog\ \langle [Q] \rangle$ holds. Expectations are therefore the quantitative analogue to predicates for classical programs.

### 3.2 Expectation Transformer for Probabilistic Programs

Let $P$ and $Q$ be probabilistic programs, $g$ a post-expectation, $x$ a program variable, $E$ an expression, $G$ a Boolean expression, and $p \in (0, 1)$. Define the *expec-* *tation transformer* $wp(\,\cdot\,,g)$ as follows [24].

$$
\begin{aligned}
wp(\mathsf{skip},g) &= g \\
wp(\mathsf{abort},g) &= 0 \\
wp(x \coloneqq E, g) &= g[x/E] \\
wp(P;Q,g) &= wp(P, wp(Q,g)) \\
wp(\mathsf{if}\,(G)\,\mathsf{then}\,\{P\}\,\mathsf{else}\,\{Q\}, g) &= [G] \cdot wp(P,g) + [\neg G] \cdot wp(Q,g) \\
wp(P[p]Q, g) &= p \cdot wp(P,g) + (1-p) \cdot wp(Q,g) \\
wp(\mathsf{while}\,(G)\,\{P\}, g) &= \mu X.([G] \cdot wp(P,X) + [\neg G] \cdot g).
\end{aligned}
$$

Here $g[x/E]$ denotes the formula obtained from $g$ by replacing free occurrences of $x$ by $E$. The least fixed point operator $\mu$ is defined over the domain of expectations [16]. It can be shown that $\langle f \rangle\, P\, \langle g \rangle$ if and only if $f \leq wp(P,g)$. That is, $wp(P,g)$ is the greatest lower bound of pre-expectation of $P$ with respect to $g$. We say $wp(P,g)$ is the *weakest pre-expectation of $P$ with respect to $g$*.

*Example 2.* The weakest pre-expectation of command $x \coloneqq x + 1\,[p]\,x \coloneqq x - 1$ with respect to $x$ is computed below:

$$
\begin{aligned}
&wp(x \coloneqq x + 1\,[p]\,x \coloneqq x - 1, x) \\
&= p \cdot wp(x \coloneqq x + 1, x) + (1-p) \cdot wp(x \coloneqq x - 1, x) \\
&= p \cdot (x+1) + (1-p) \cdot (x-1) \\
&= x + 2p - 1
\end{aligned}
$$

It follows that $\langle x + 2p - 1 \rangle\, x \coloneqq x + 1\,[p]\,x \coloneqq x - 1\, \langle x \rangle$ holds.

## 4 Quantitative Loop Invariants

Given a pre-expectation $preE$, a post-expectation $postE$, a Boolean expression $G$, and a loop-free probabilistic program $body$, we would like to verify whether

$$\langle preE \rangle\, \mathsf{while}\,(G)\,\{body\}\, \langle postE \rangle$$

holds or not. One way to solve this problem is to compute the weakest pre-expectation $wp(\mathsf{while}\,(G)\,\{body\}, postE)$ and check if it is not less than $preE$ pointwisely. However, the weakest pre-expectation of a while-command requires fixed point computation. To avoid the expensive computation, we can solve the problem by finding quantitative loop invariants.

**Theorem 1 ([15, 24]).** *Let $preE$ be a pre-expectation, $postE$ a post-expectation, $G$ a Boolean expression, and $body$ a loop-free probabilistic program. To show*

$$\langle preE \rangle\, \mathsf{while}\,(G)\,\{body\}\, \langle postE \rangle,$$

*it suffices to find a* loop invariant *$I$ which is an expectation such that*

1. *(boundary)* $preE \leq I$ *and* $I \cdot [\neg G] \leq postE$;

2. *(invariant)* $I \cdot [G] \leq wp(body, I)$;
3. *(soundness)* the loop terminates from any state in $G$ with probability $1$, and
   (a) the number of iterations is finite;
   (b) $I$ is bounded above by some fixed constant; or
   (c) the expected value of $I \cdot [G]$ tends to zero as the loop continues to iterate.

In this paper, we only focus on checking the boundary and invariant conditions in Theorem 1. One however can show that *any* polynomial expectation is sound for all examples we consider. In fact, one can establish the soundness of a large class of loop invariants before any specific invariant is found. For instance, it can be shown that any polynomial expectation satisfies the third soundness condition, as long as the probability of exiting the loop is bounded below by a non-zero constant in each iteration. We refer the reader to [24] for more details of sufficient conditions for soundness.

## 5 Multivariate Lagrange Interpolation

Fix a degree $n$ of quantitative loop invariants and number of variables $m$. Let $d = \binom{m+n}{n}$. *Multivariate Lagrange interpolation* is a method to construct an explicit expression for any polynomial in $\mathbb{R}[\mathbf{x}_m^n]$ by sampling, see e.g., [6, 26, 30]. Given $d$ sampling points $\mathbf{s}_1, \mathbf{s}_2, \ldots, \mathbf{s}_d \in \mathbb{R}^m$, we can compute a Lagrange basis as follows [28]. Let $\{b_1, b_2, \ldots, b_d\} = \{x_1^{d_1} x_2^{d_2} \cdots x_m^{d_m} : d_1 + d_2 + \cdots + d_m \leq n\}$ be the set of monomials in $\mathbb{R}[\mathbf{x}_m^n]$. For $1 \leq i \leq d$, define

$$M_i = det \begin{bmatrix} b_1(\mathbf{s}_1) & \cdots & b_d(\mathbf{s}_1) \\ \vdots & & \vdots \\ b_1 & \cdots & b_d \\ \vdots & & \vdots \\ b_1(\mathbf{s}_d) & \cdots & b_d(\mathbf{s}_d) \end{bmatrix} \leftarrow \text{the } i\text{th row}$$

Observe that $M_i \in \mathbb{R}[\mathbf{x}_m^n]$ for $1 \leq i \leq d$. Moreover, $M_i(\mathbf{s}_j) = 0$ for $i \neq j$ and $M_1(\mathbf{s}_1) = M_2(\mathbf{s}_2) = \cdots = M_d(\mathbf{s}_d) = r$ for some $r \in \mathbb{R}$. If $r = 0$, then there is a geometrical dependency among the sampling points $\mathbf{s}_1, \mathbf{s}_2, \ldots, \mathbf{s}_d$ [2], and thus no Lagrange basis could be determined from these points. If $r \neq 0$, define $B_i = M_i/r$ for $1 \leq i \leq d$. Then $\mathcal{B}(\mathbf{s}_1, \mathbf{s}_2, \ldots, \mathbf{s}_d) = \{B_i : 1 \leq i \leq d\} \subseteq \mathbb{R}[\mathbf{x}_m^n]$ is called a *Lagrange basis* of $\mathbb{R}[\mathbf{x}_m^n]$.

Observe that $B_i(\mathbf{s}_j) = [i = j]$ for $1 \leq i, j \leq d$. Thus $\sum_{i=1}^{d} f(\mathbf{s}_i) B_i(\mathbf{s}_j) = f(\mathbf{s}_j)$ for $1 \leq j \leq d$. Moreover, given any $f \in \mathbb{R}[\mathbf{x}_m^n]$, we can write $f = \sum_{i=1}^{d} f(\mathbf{s}_i) B_i$. Define the *Lagrange functional* $\mathcal{L}[\mathbf{s}_1, \mathbf{s}_2, \ldots, \mathbf{s}_d] : \mathbb{R}^d \to \mathbb{R}[\mathbf{x}_m^n]$ by

$$\mathcal{L}[\mathbf{s}_1, \mathbf{s}_2, \ldots, \mathbf{s}_d](c_1, c_2, \ldots, c_d) = \sum_{i=1}^{d} c_i B_i.$$

Then $f = \mathcal{L}[\mathbf{s}_1, \mathbf{s}_2, \ldots, \mathbf{s}_d](f(\mathbf{s}_1), f(\mathbf{s}_2), \ldots, f(\mathbf{s}_d))$ for any $f \in \mathbb{R}[\mathbf{x}_m^n]$. We shall call $f(\mathbf{s}_1), f(\mathbf{s}_2), \ldots, f(\mathbf{s}_d) \in \mathbb{R}$ the *coefficients* for $f$ on basis $\mathcal{B}(\mathbf{s}_1, \mathbf{s}_2, \ldots, \mathbf{s}_d)$.

## 6 Interpolation of Loop Invariants

Suppose we would like to find a quantitative loop invariant $I \in \mathbb{R}[\mathbf{x}_m^n]$ for

$$\langle preE \rangle \text{ while } (G) \{body\} \langle postE \rangle$$

where $preE$ is a pre-expectation, $postE$ is a post-expectation, $G$ is a Boolean expression, and $body$ is a loop-free probabilistic program. Assume the soundness of $I$ can be verified. We shall use Lagrange interpolation to find $I$.

Let $\mathbf{s}_1, \mathbf{s}_2, \ldots, \mathbf{s}_d \in \mathbb{R}^m$ be sampling points that determine a Lagrange basis. If the coefficients $I(\mathbf{s}_1), I(\mathbf{s}_2), \ldots, I(\mathbf{s}_d) \in \mathbb{R}$ are known, then

$$I = \mathcal{L}[\mathbf{s}_1, \mathbf{s}_2, \ldots, \mathbf{s}_d](I(\mathbf{s}_1), I(\mathbf{s}_2), \ldots, I(\mathbf{s}_d))$$

by Lagrange interpolation. Our idea therefore is to find the coefficients via constraint-solving. By the boundary and invariant conditions in Theorem 1, we have the following requirements about any loop invariant $I$:

$$\begin{aligned} preE &\leq I \\ I \cdot [\neg G] &\leq postE \\ I \cdot [G] &\leq wp(body, I). \end{aligned} \qquad (1)$$

*Example 3.* Consider

$$\langle xy - x^2 \rangle \; z := 0; \text{ while } (0 < x < y) \, \{ \, x := x+1 \, [0.5] \, x := x-1; \; z := z+1; \, \} \, \langle z \rangle.$$

The following must hold for any loop invariant $I$

$$\begin{aligned} xy - x^2 &\leq I \\ I \cdot [x \leq 0 \vee y \leq x] &\leq z \\ I \cdot [0 < x < y] &\leq 0.5 \cdot I(x+1, y, z+1) + 0.5 \cdot I(x-1, y, z+1). \end{aligned}$$

Observe that $wp(x := x+1 \, [0.5] \, x := x-1; \; z := z+1, I(x,y,z)) = wp(x := x+1 \, [0.5] \, x := x-1, I(x,y,z+1)) = 0.5 \cdot wp(x := x+1, I(x,y,z+1)) + 0.5 \cdot wp(x := x-1, I(x,y,z+1)) = 0.5 \cdot I(x+1, y, z+1) + 0.5 \cdot I(x-1, y, z+1)$.

Requirements (1) can have indicators on both sides of inequality, which is beyond the capability of the solvers we use. We would like to obtain a constraint by removing indicators in two steps. First, we rewrite the expectations to a normal form. An expectation is in *disjoint normal form (DNF)* if it is of the form $f = [P_1] \cdot f_1 + \cdots + [P_k] \cdot f_k$, where $P_1, P_2, \ldots, P_k$ are disjoint, that is, at most one of $P_1, P_2, \ldots, P_k$ evaluates to true on any valuation.

**Theorem 2 ([22]).** *Given an expectation of the form $f = [P_1] \cdot f_1 + \cdots + [P_k] \cdot f_k$, $f$ is equivalent to the following expectation in DNF:*

$$\sum_{I \subseteq K} \left[ \left( \bigwedge_{i \in I} P_i \right) \wedge \neg \left( \bigwedge_{j \in K \setminus I} P_j \right) \right] \cdot \sum_{i \in I} f_i$$

*where $K = \{1, 2, \ldots, k\}$.*

We then transform inequalities between expectations in DNF to constraints.

**Theorem 3 ([15]).** *Suppose $f = [P_1] \cdot f_1 + \cdots + [P_k] \cdot f_k$ and $g = [Q_1] \cdot g_1 + \cdots + [Q_h] \cdot g_h$ are expectations over $\mathbf{x}_m$ in DNF. $f \leq g$ iff for every $\mathbf{x}_m$*

$$\bigwedge_{j \in K} \bigwedge_{i \in H} ((P_j \wedge Q_i) \Rightarrow f_j \leq g_i) \quad \wedge$$

$$\bigwedge_{j \in K} \left( \left( \bigwedge_{i \in H} \neg Q_i \wedge P_j \right) \Rightarrow f_j \leq 0 \right) \wedge \bigwedge_{i \in H} \left( \left( \bigwedge_{j \in K} \neg P_j \wedge Q_i \right) \Rightarrow 0 \leq g_i \right)$$

*where $K = \{1, 2, \ldots, k\}$ and $H = \{1, 2, \ldots, h\}$.*

*Example 4.* By Theorem 2 and 3, requirements in Example 3 are equivalent to

$$xy - x^2 \leq I \quad \wedge$$
$$(x \leq 0 \vee y \leq x) \Rightarrow I \leq z \quad \wedge$$
$$(x \leq 0 \vee y \leq x) \Rightarrow 0 \leq z \quad \wedge$$
$$(0 < x < y) \Rightarrow I \leq 0.5 \cdot I(x+1, y, z+1) + 0.5 \cdot I(x-1, y, z+1) \quad \wedge$$
$$(0 < x < y) \Rightarrow 0 \leq 0.5 \cdot I(x+1, y, z+1) + 0.5 \cdot I(x-1, y, z+1)$$

for every $x, y, z$.

We define the *loop invariant constraint* $\phi[\mathbf{s}_1, \mathbf{s}_2, \ldots, \mathbf{s}_d](c_1, c_2, \ldots, c_d)$ as the constraint transformed from the requirements (1), where the quantitative loop invariant $I$ is replaced by Lagrange functional $\mathcal{L}[\mathbf{s}_1, \mathbf{s}_2, \ldots, \mathbf{s}_d](c_1, c_2, \ldots, c_d)$.

*Example 5.* We have the following loop invariant constraint from Example 4.

$$\phi[\mathbf{s}_1, \mathbf{s}_2, \ldots, \mathbf{s}_{10}](c_1, c_2, \ldots, c_{10}) =$$
$$xy - x^2 \leq \mathcal{L}[\mathbf{s}_1, \mathbf{s}_2, \ldots, \mathbf{s}_{10}](c_1, c_2, \ldots, c_{10}) \quad \wedge$$
$$(x \leq 0 \vee y \leq x) \Rightarrow \mathcal{L}[\mathbf{s}_1, \mathbf{s}_2, \ldots, \mathbf{s}_{10}](c_1, c_2, \ldots, c_{10}) \leq z \quad \wedge$$
$$(0 < x < y) \Rightarrow 2 \cdot \mathcal{L}[\mathbf{s}_1, \mathbf{s}_2, \ldots, \mathbf{s}_{10}](c_1, c_2, \ldots, c_{10}) \leq$$
$$\mathcal{L}[\mathbf{s}_1, \mathbf{s}_2, \ldots, \mathbf{s}_{10}](c_1, c_2, \ldots, c_{10})(x+1, y, z+1) +$$
$$\mathcal{L}[\mathbf{s}_1, \mathbf{s}_2, \ldots, \mathbf{s}_{10}](c_1, c_2, \ldots, c_{10})(x-1, y, z+1).$$

With loop invariant constraints, it is easy to state our goal. Observe that $\exists \mathbf{s}_1, \mathbf{s}_2, \ldots, \mathbf{s}_d \, \exists c_1, c_2, \ldots, c_d \, \forall \mathbf{x}_m. \, \phi[\mathbf{s}_1, \mathbf{s}_2, \ldots, \mathbf{s}_d](c_1, c_2, \ldots, c_d)$ implies the existence of a quantitative loop invariant satisfying the boundary and invariant conditions in Theorem 1. Our strategy hence is to choose sampling points $\mathbf{s}_1, \mathbf{s}_2, \ldots, \mathbf{s}_d$ such that $\exists c_1, c_2, \ldots, c_d \, \forall \mathbf{x}_m. \, \phi[\mathbf{s}_1, \mathbf{s}_2, \ldots, \mathbf{s}_d](c_1, c_2, \ldots, c_d)$ holds.

We will choose sampling points to simplify the loop invariant constraint. Recall that sampling points are not unique in Lagrange interpolation. For a loop invariant constraint, we select sampling points so that several coefficients among $c_1, c_2, \ldots, c_d$ are determined. This helps us to evaluate the quantified loop invariant constraint $\exists c_1, c_2, \ldots, c_d \, \forall \mathbf{x}_m. \, \phi[\mathbf{s}_1, \mathbf{s}_2, \ldots, \mathbf{s}_d](c_1, c_2, \ldots, c_d)$.

To evaluate the quantified loop invariant constraint, observe that the Lagrange functional $\mathcal{L}[\mathbf{s}_1, \mathbf{s}_2, \ldots, \mathbf{s}_d](c_1, c_2, \ldots, c_d)$ is a multivariate polynomial over $c_1, c_2, \ldots, c_d$ and $\mathbf{x}_m$. A loop invariant constraint is hence non-linear. However, $\phi[\mathbf{s}_1, \mathbf{s}_2, \ldots, \mathbf{s}_d](c_1, c_2, \ldots, c_d)(\mathbf{e})$ is a linear constraint over coefficients for every *experiment* $\mathbf{e} \in \mathbb{Z}^m$, i.e., valuation over $\mathbf{x}_m$. We therefore use experiments to construct a series of linear constraints and find coefficients by an SMT solver.

**Input**: $\langle preE \rangle$ while $(G)$ $\{body\}$ $\langle postE \rangle$ : a loop over program variables $\mathbf{x}_m$; $n$ : the degree of an loop invariant
**Output**: $I$ : a loop invariant satisfying the boundary and invariant conditions in Theorem 1

$d \leftarrow \binom{m+n}{n}$;
$\mathbf{s}_1, \mathbf{s}_2, \ldots, \mathbf{s}_d \leftarrow$ SamplingPoints();
$C \leftarrow$ InitialConstraint$(\mathbf{s}_1, \mathbf{s}_2, \ldots, \mathbf{s}_d)$;
**while** *C has a model* **do**
$\quad \hat{c}_1, \hat{c}_2, \ldots, \hat{c}_d \leftarrow$ a model of $C$ from an SMT solver;
$\quad$ **switch** *RandomExperiments*$(C)$ **do**
$\quad\quad$ **case** *Pass*:
$\quad\quad\quad$ **switch** *UQElem*$(\mathbf{x}_m, \phi[\mathbf{s}_1, \mathbf{s}_2, \ldots, \mathbf{s}_d](\hat{c}_1, \hat{c}_2, \ldots, \hat{c}_d))$ **do**
$\quad\quad\quad\quad$ **case** *True*: **return** $\mathcal{L}[\mathbf{s}_1, \mathbf{s}_2, \ldots, \mathbf{s}_d](\hat{c}_1, \hat{c}_2, \ldots, \hat{c}_d)$ ;
$\quad\quad\quad\quad$ **case** *CounterExample* $(\mathbf{e})$ : RefineConstraint$(C, \mathbf{e})$ ;
$\quad\quad\quad$ **endsw**
$\quad\quad$ **case** *CounterExample* $(\mathbf{e})$ : RefineConstraint$(C, \mathbf{e})$ ;
$\quad$ **endsw**
**end**
// No loop invariant

**Algorithm 1:** Quantitative loop invariant synthesis

Algorithm 1 shows our top-level algorithm. The algorithm starts by choosing sampling points (Section 6.1). The sampling points are then used to construct the initial linear constraint over coefficients (Section 6.2). The while loop evaluates the quantified loop invariant constraint $\exists c_1, c_2, \ldots, c_d \, \forall \mathbf{x}_m. \, \phi[\mathbf{s}_1, \mathbf{s}_2, \ldots, \mathbf{s}_d](c_1, c_2, \ldots, c_d)$. In each iteration, the algorithm selects coefficients $\hat{c}_1, \hat{c}_2, \ldots, \hat{c}_d$ by a model of the linear constraint obtained from an SMT solver. It then checks whether $\forall \mathbf{x}_m. \, \phi[\mathbf{s}_1, \mathbf{s}_2, \ldots, \mathbf{s}_d](\hat{c}_1, \hat{c}_2, \ldots, \hat{c}_d)$ is true. The algorithm does this by first trying a number of random experiments (Section 6.3). Only after the random experiments are passed, will the algorithm performs universal quantifier elimination to evaluate the quantified constraint (Section 6.4). If the random experiments fail, or quantifier elimination does not evaluate to true, our algorithm refines the linear constraint by a counterexample and reiterates (Section 6.5).

### 6.1 Choosing Sampling Points

In Lagrange interpolation, sampling points need be chosen in the first place. We would like to choose sampling points so that as many coefficients are determined as possible. To this end, observe that $\mathcal{L}[\mathbf{s}_1, \mathbf{s}_2, \ldots, \mathbf{s}_d](c_1, c_2, \ldots, c_d)(\mathbf{s}_i) = c_i$ for $1 \le i \le d$. In other words, $\phi[\mathbf{s}_1, \mathbf{s}_2, \ldots, \mathbf{s}_d](c_1, c_2, \ldots, c_d)(\mathbf{s}_i)$ can be significantly simplified if a sampling point $\mathbf{s}_i$ is used as an experiment. Consider, for instance, the boundary condition in our running example:

$$xy - x^2 \leq \mathcal{L}[\mathbf{s}_1, \mathbf{s}_2, \ldots, \mathbf{s}_d](c_1, c_2, \ldots, c_d)(x, y, z); \text{ and}$$
$$(x \leq 0 \vee y \leq x) \Rightarrow \mathcal{L}[\mathbf{s}_1, \mathbf{s}_2, \ldots, \mathbf{s}_d](c_1, c_2, \ldots, c_d)(x, y, z) \leq z$$

If $\mathbf{s}_j = (0, 3, 0)$ is a sampling point, then the condition can be simplified to $0 \leq c_j$ and $c_j \leq 0$. Thus $c_j$ is determined by choosing $(0, 3, 0)$ as both a sampling point and an experiment.

Ideally, one would choose sampling points so that all coefficients are determined. Unfortunately, such points tend to be geometrically dependent in practice. Thus we cannot expect to establish a Lagrange basis from these points exclusively. Instead, we try to find sampling points which yield a Lagrange basis and determine as many coefficients as possible. We adopt a weighted random search for this purpose. That is, we pick sampling points randomly according to their weights, so that points determining more coefficients are more likely to be picked. If the randomly selected sampling points fail to yield a Lagrange basis, we discard them and select other sampling points randomly again. In our experiments, this heuristic finds pretty good sampling points in reasonable time.

## 6.2 Initial Constraint

After sampling points are chosen, we compute the initial linear constraint over coefficients. Recall that $\mathcal{L}[\mathbf{s}_1, \mathbf{s}_2, \ldots, \mathbf{s}_d](c_1, c_2, \ldots, c_d)(\mathbf{s}_i) = c_i$ for $1 \leq i \leq d$. By taking sampling points as experiments, the loop invariant constraint $\phi[\mathbf{s}_1, \mathbf{s}_2, \ldots, \mathbf{s}_d](c_1, c_2, \ldots, c_d)$ is simplified to a linear constraint over $c_1, c_2, \ldots, c_d$.

*Example 6.* Consider the loop invariant constraint in Example 5. We first choose 10 sampling points $\mathbf{s}_1, \ldots, \mathbf{s}_{10}$ (see table below) to establish a Lagrange basis. We then compute the initial constraints by simplifying the loop invariant constraint with the sampling points. For example, we obtain constraint $c_2 = 0$ from point $\mathbf{s}_2 = (2, 2, 0)$ as follows:

$\phi[\mathbf{s}_1, \mathbf{s}_2, \ldots, \mathbf{s}_{10}](c_1, c_2, \ldots, c_{10})(2, 2, 0)$
iff $(2 \cdot (2 - 2) \leq c_2) \wedge ((2 \leq 0 \vee 2 \leq 2) \Rightarrow c_2 \leq 0) \wedge$
$\quad (0 < 2 < 2 \Rightarrow 0 \leq -2c_1 - 42c_2 + 3c_3 + 27c_4 + 9c_5 + 6c_6 + 14c_7 - 12c_8 - 3c_{10})$
iff $0 \leq c_2 \wedge c_2 \leq 0$ iff $c_2 = 0$.

We list all initial constraints in the following table, where $\phi[\mathbf{s}_1, \mathbf{s}_2, \ldots, \mathbf{s}_{10}](c_1, c_2, \ldots, c_{10})(\mathbf{s}_i)$ is denoted by $\psi(\mathbf{s}_i)$ for simplicity.

| $i$ | $\mathbf{s}_i$ | $\psi(\mathbf{s}_i)$ | $i$ | $\mathbf{s}_i$ | $\psi(\mathbf{s}_i)$ | $i$ | $\mathbf{s}_i$ | $\psi(\mathbf{s}_i)$ |
|---|---|---|---|---|---|---|---|---|
| 1 | $0, 3, 0$ | $c_1 = 0$ | 2 | $2, 2, 0$ | $c_2 = 0$ | 3 | $0, 3, 1$ | $0 \leq c_3 \leq 1$ |
| 4 | $1, 1, 0$ | $c_4 = 0$ | 5 | $1, 1, 2$ | $0 \leq c_5 \leq 2$ | 6 | $2, 2, 1$ | $0 \leq c_6 \leq 1$ |
| 7 | $3, 3, 0$ | $c_7 = 0$ | 8 | $0, 0, 1$ | $0 \leq c_8 \leq 1$ | 9 | $0, 1, 0$ | $c_9 = 0$ |
| 10 | $2, 3, 3$ | $6 \leq 3c_{10} \leq -4c_1 - 36c_2 + 5c_3 + 30c_4 + 12c_5 - 6c_6 + 16c_7 - 14c_8$ | | | | | | |

Note that our choice of sampling points helps the initial constraints determine 5 coefficients. If a standard monomial basis were used, none of the coefficients could be determined by the initial constraints.

### 6.3 Random Experiments

From a linear constraint of coefficients, we obtain a model $\hat{c}_1, \hat{c}_2, \ldots, \hat{c}_d$ of the linear constraint from an SMT solver. Recall that we would like to check if $\forall \mathbf{x}_m. \phi[\mathbf{s}_1, \mathbf{s}_2, \ldots, \mathbf{s}_d](\hat{c}_1, \hat{c}_2, \ldots, \hat{c}_d)$ is true. Before using expensive quantifier elimination immediately, we first perform a number of random tests. If $\phi[\mathbf{s}_1, \mathbf{s}_2, \ldots, \mathbf{s}_d]$ $(\hat{c}_1, \hat{c}_2, \ldots, \hat{c}_d)(\mathbf{e})$ evaluates to true for all random experiments $\mathbf{e} \in \mathbb{Z}^m$, the coefficients $\hat{c}_1, \hat{c}_2, \ldots, \hat{c}_d$ may induce a loop invariant. Otherwise, a witness experiment $\mathbf{e}$ is used to refine the linear constraint over coefficients.

When the coefficients do not induce a loop invariant, the random experiments make it possible to avoid expensive quantifier elimination and to obtain a witness experiment without resorting to an SMT solver. This possibility is important, because the solver we use does not always find a valid witness experiment.

### 6.4 Universal Quantifier Elimination

After random tests, we perform quantifier elimination check if $\forall \mathbf{x}_m. \phi[\mathbf{s}_1, \mathbf{s}_2, \ldots, \mathbf{s}_d](\hat{c}_1, \hat{c}_2, \ldots, \hat{c}_d)$ is true. If so, the polynomial $\mathcal{L}[\mathbf{s}_1, \mathbf{s}_2, \ldots, \mathbf{s}_d]$ $(\hat{c}_1, \hat{c}_2, \ldots, \hat{c}_d)$ is a quantitative loop invariant satisfying the boundary and invariant conditions. Otherwise, we obtain a witness experiment to refine our linear constraint.

Universal quantifier elimination is carried out in two steps. We first eliminate the quantifiers in the ordered field theory [3, 11]. Intuitively, the ordered field theory formalizes real numbers $\mathbb{R}$. Since quantifier elimination tools such as REDLOG [10] employ algebra and real algebraic geometry, eliminating quantifiers over real numbers is more efficient than over integers. If $\forall \mathbf{x}_m. \phi[\mathbf{s}_1, \mathbf{s}_2, \ldots, \mathbf{s}_d]$ $(\hat{c}_1, \hat{c}_2, \ldots, \hat{c}_d)$ is true over $\mathbb{R}$, it is also true over $\mathbb{Z}$. Thus $\hat{c}_1, \hat{c}_2, \ldots, \hat{c}_d$ induces a quantitative loop invariant. Otherwise, we perform quantifier elimination over $\mathbb{Z}$.

If $\forall \mathbf{x}_m. \phi[\mathbf{s}_1, \mathbf{s}_2, \ldots, \mathbf{s}_d](\hat{c}_1, \hat{c}_2, \ldots, \hat{c}_d)$ evaluates to true over $\mathbb{Z}$, we are done. Otherwise, quantifier elimination gives a constraint equivalent to the quantified query. We then use an SMT solver to obtain a witness experiment. We abort the procedure if the solver times-out or fails to yield a valid witness experiment.

### 6.5 Constraint Refinement

Let $\mathbf{e} = (\hat{x}_1, \hat{x}_2, \ldots, \hat{x}_m) \in \mathbb{Z}^m$ be a witness experiment such that $\phi[\mathbf{s}_1, \mathbf{s}_2, \ldots, \mathbf{s}_d]$ $(\hat{c}_1, \hat{c}_2, \ldots, \hat{c}_d)(\mathbf{e})$ evaluates to false. Recall that we would like to find coefficients $c_1, c_2, \ldots, c_d$ such that $\phi[\mathbf{s}_1, \mathbf{s}_2, \ldots, \mathbf{s}_d](c_1, c_2, \ldots, c_d)$ is true for every valuations over $\mathbf{x}_m$. Particularly, $\phi[\mathbf{s}_1, \mathbf{s}_2, \ldots, \mathbf{s}_d](c_1, c_2, \ldots, c_d)(\hat{x}_1, \hat{x}_2, \ldots, \hat{x}_m)$ must also be true for such coefficients. Note that $\phi[\mathbf{s}_1, \mathbf{s}_2, \ldots, \mathbf{s}_d](c_1, c_2, \ldots, c_d)(\hat{x}_1, \hat{x}_2, \ldots, \hat{x}_m)$ is a linear constraint on coefficients $c_1, c_2, \ldots, c_d$ that excludes the incorrect coefficients $\hat{c}_1, \hat{c}_2, \ldots, \hat{c}_d$. By adding the linear constraint to the current set of constraints, our algorithm will find different coefficients in the next iteration.

# 7  Applications

| Name | preE | postE | Time | TO | L | T | S | #L | #T | #S |
|---|---|---|---|---|---|---|---|---|---|---|
| ruin | $xy - x^2$ | $z$ | 3.6s | 0% | 0.3s | 2.8s | 0.3s | 5.2 | 61.5 | 5.0 |
| geo1 | $x + 3zy$ | $x$ | 3.0s | 0% | 1.4s | 1.5s | 0.1s | 21.4 | 32.1 | 1.0 |
| geo2 | $x + \frac{15}{2}z$ | $x$ | 8.0s | 0% | 1.4s | 4.9s | 0.2s | 22.3 | 108 | 3.8 |
| bin1 | $x + \frac{1}{4}ny$ | $x$ | 4.5s | 0% | 1.4s | 2.9s | 0.1s | 22.7 | 64.0 | 1.0 |
| bin2 | $\frac{1}{8}n^2 - \frac{1}{8}n + \frac{3}{4}ny$ | $x$ | 77.5s | 19% | 0.4s | 9.0s | 15.8s | 5.9 | 185 | 10.3 |
| sum | $\frac{1}{4}n^2 + \frac{1}{4}n$ | $x$ | 2.5s | 0% | 0.1s | 1.9s | 0.4s | 1.2 | 42.7 | 5.8 |
| prod | $\frac{1}{4}n^2 - \frac{1}{4}n$ | $xy$ | 15.7s | 5% | 0.3s | 4.3s | 2.3s | 4.2 | 97.0 | 6.5 |
| coin1 | $\frac{1}{2} - \frac{1}{2}x$ | $1 - x + xy$ | 2.0s | 0% | 0.4s | 1.2s | 0.3s | 5.7 | 27.1 | 3.6 |
| coin2 | $\frac{1}{2} - \frac{1}{2}y$ | $x + xy$ | 3.9s | 0% | 0.6s | 1.4s | 1.8s | 9.8 | 32.0 | 4.2 |
| coin3 | $\frac{8}{3} - \frac{8}{3}x - \frac{8}{3}y + \frac{1}{3}n$ | $n$ | 18.5s | 1% | 12.6s | 2.6s | 1.4s | 202 | 57.9 | 7.2 |

**Table 1.** Summary of results. The name of each experiment is shown in column Name. The annotated pre- and post-expectations are shown in columns preE and postE, respectively. Column Time lists the mean execution times our prototype took to verify the annotations, and TO denotes the timeout ratios. Besides, columns L, T, and S show the average times our prototype spent in sampling a Lagrange basis, making random tests and synthesizing coefficients, respectively. Finally, columns #L, #T, and #S show the average numbers of iterations our prototype has taken to find sampling points, make random tests, and refine constraints, respectively. The last six columns are calculated based on the runs that finished within timeouts.

We have implemented a prototype in JavaScript to test our techniques. For each simple loop, we manually perform the weakest pre-expectation computation and the DNF transformation to translate the requirements (1) into loop invariant constraints. We then use our prototype to find a quantitative loop invariant based on the constraints. Our prototype uses GNU OCTAVE [12] to compute Lagrange interpolation, Z3 [7] to solve linear constraints, and REDLOG [10] to perform quantifier elimination. The experiments are done on an Intel Xeon 3.07GHz Linux workstation with 16GB RAM.

We consider six types of applications: gambler's ruin problem, geometric distribution, binomial distribution, sum of random series, product of random variables, and simulation of a fair coin. We also consider variants of geometric and binomial distributions. For the fair-coin simulation, we find three quantitative loop invariants to prove the correctness and the expected execution time of the simulation. In each probabilistic program, we annotate the while loop with a pre-expectation and a post-expectation. Note that the annotated pre-expectation is by construction a precise estimate of the annotated post-expectation.

Our results are summarized in Table 1. We use a fixed random seed for all experiments and calculate the averages over 1000 runs with a 300-second timeout. Due to random sampling, the prototype may synthesize different loop invariants in different runs of the same experiment. We now discuss the applications in more details.

**Gambler's Ruin Problem.** In Example 1, we consider a game where a player has $x$ dollars initially and plays until he loses all his money or wins up to $y - x$ dollars for some $y > x$. The expected number of rounds before the game ends is $E[z] = x \cdot (y-x)$. Our prototype proves this result within 3.6 seconds on average.

**Geometric Distribution.** The geometric distribution describes the number of tails before the first head in a sequence of coin-tossing. When the probability of head is 0.25, we expect to see $\frac{1-0.25}{0.25} = 3$ tails before the first head. The following program computes a geometrically distributed random variable $x$:

$$x := 0; \ z := 1; \ \mathsf{while} \ (z \neq 0) \ \{ \ z := 0 \ [0.25] \ x := x + y; \ \}$$

Our prototype finds a quantitative loop invariant for the pre-expectation $E[x] = 3y$ within 3 seconds on average.

We moreover consider the following variant of the game. A player keeps flipping a coin until the head turns up. He wins $k$ dollars if the tail turns up at the $k$th flip. The variant is modeled as follows.

$$x := 0; \ y := 0; \ z := 1; \ \mathsf{while} \ (z \neq 0) \ \{ \ y := y + 1; \ z := 0 \ [0.25] \ x := x + y; \ \}$$

The expected amount of money a player can win is $E[x] = \frac{1}{2} \left( 0.25^{-2} - 1 \right) = \frac{15}{2}$. Our prototype proves this result within 8 seconds on average.

**Binomial Distribution.** The binomials distribution describes the number of heads that appear in a fixed number of coin-tossing. If the probability of head is 0.25 and the number of tosses is $n$, then the expected number of heads is $0.25n$. The following program computes a binomially distributed random variable $x$:

$$x := 0; \ \mathsf{while} \ (0 < n) \ \{ \ x := x + y \ [0.25] \ \mathsf{skip}; \ n := n - 1; \ \}$$

Our prototype proves $E[x] = 0.25ny$ within 4.5 seconds on average. We moreover consider the following variant. A player flips a coin for $n$ times. At the $k$th flip, he wins $k$ dollars if the head turns up and wins $y$ dollars otherwise. This game can be modeled as follows.

$$x := 0; \ \mathsf{while} \ (0 < n) \ \{ \ x := x + n \ [0.25] \ x := x + y; \ n := n - 1; \ \}$$

The expected amount of money a player can win is $E[x] = 0.25 \cdot \frac{1}{2}n(n+1) + (1 - 0.25) \cdot ny = \frac{1}{8}n^2 - \frac{1}{8}n + \frac{3}{4}ny$. Our prototype proves this result within 77.5 seconds on average.

**Sum of Random Series.** Consider a game where a player flips a coin for $n$ times. The player wins $k$ dollars if the head turns up at the $k$th flip. The following program models this game when the head probability of the coin is 0.5:

$$x := 0; \ \mathsf{while} \ (0 < n) \ \{ \ x := x + n \ [0.5] \ \mathsf{skip}; \ n := n - 1; \ \}$$

The expected amount of money the player can win from this game is $E[x] = 0.5 \cdot \sum_{i=1}^{n} i = 0.5 \cdot \frac{1}{2}n(n+1)$ dollars. Our prototype proves this result within 2.5 seconds on average.

**Product of Dependent Random Variables.** We consider a game where two players flip a coin for $n$ times. The first player wins one dollar for each head and the second player wins one dollars for each tail. When the head probability of the coin is 0.5, this game can be modeled by the following program where variables $x, y$ represent the amount of money won by the respective players:

$$\mathsf{while}\,(0 < n)\,\{\ x \coloneqq x + 1\,[0.5]\,y \coloneqq y + 1;\ n \coloneqq n - 1;\ \}$$

It can be shown that $E[xy] = \frac{1}{4}(n^2 - n)$. Our prototype proves this result within 15.7 seconds on average.

**Simulation of a Fair Coin.** We consider an algorithm that simulates a fair coin flip using biased coins [15]:

$$x \coloneqq 0;\ y \coloneqq 0;\ n \coloneqq 0;$$
$$\mathsf{while}\,(x = y)\,\{\ x \coloneqq 1\,[0.25]\,x \coloneqq 0;\ y \coloneqq 1\,[0.25]\,y \coloneqq 0;\ n \coloneqq n + 1;\ \}$$

The algorithm uses two biased coins $x$ and $y$ with head probability 0.25. The main loop flips the two coins at each iteration and terminates when the coins show different outcomes. The value of $x$ is then taken as the final outcome, with 1 representing the head and 0 representing the tail.

To show that the algorithm indeed simulates a fair coin flip, we prove

$$0.5 - 0.5x \le wp(loop, 1 - x + xy) \quad \text{and} \quad 0.5 - 0.5y \le wp(loop, x + xy),$$

where *loop* denotes the while loop in the program. Since $x = y = 0$ before the loop starts and $xy = 0$ after the loop stops, we see that $0.5 \le E[1-x]$ and $0.5 \le E[x]$ on termination. Since $x \in \{0, 1\}$, it follows that $\Pr\{x = 1\} = \Pr\{x = 0\} = 0.5$ on termination, and thus the correctness of the algorithm is concluded.

Observe moreover that the number of iterations made before termination obeys a geometric distribution with head probability $0.25 \cdot 2(1 - 0.25) = 0.375$. Hence, the expected number of iterations is $E[n] = \frac{1-0.375}{0.375} + 1 = \frac{8}{3}$. This result is verified by our prototype within 18.5 seconds on average.

## 8 Evaluation

Our technique is closely related to the PRINSYS tool [15], which implements the constraint-based quantitative invariant synthesis approach developed in [22]. PRINSYS receives a probabilistic program and a template with unknown coefficients. It derives loop invariant constraints from the template and exploits SMT-solvers to perform quantifier elimination and simplification for the constraints. The tool generates a formula, which is in effect a conjunction of non-linear inequalities, describing all coefficients that make the supplied template an inductive loop invariant. A concrete quantitative invariant has to be derived manually by extracting solutions from the formula.

For our prototype, the input is a quantitative Hoare triple and there are three possible outputs: "unknown" (due to timeout or invalid counterexamples), "disproved" with a witness (a valuation of program variables), and "proved" with a proof (a quantitative loop invariant). For PRINSYS, it receives a program and a template, and outputs a constraint describing all inductive loop invariants in form of the template. To verify a specific Hoare triple with PRINSYS, one has to encode the interested pre- and post-expectations as well as the form of possible invariants into the same template. Designing a template for PRINSYS is a tricky task that needs to be done on a case-by-case basis. In contrast, our technique does not require manually supplied templates, though the degree of loop invariants has to be fixed a priori.

One could use templates to represent non-linear loop invariants. We nevertheless failed to verify any of our non-linear examples with PRINSYS. In particular, we could not generate formulae that subsume the quantitative loop invariants computed by our prototype. This however does not imply that our examples are beyond the capability of PRINSYS, since we could not arguably try all templates manually. The designers of PRINSYS also examined their tool on some non-linear examples, e.g., the gambler's ruin problem, and reported negative results in [15]. Generally, when the supplied template is non-linear, it becomes intractable to derive a loop invariant, or even to decide the existence of a loop invariant, from the formula yielded by PRINSYS. Maybe a counterexample-refinement approach is helpful here, but this requires further research and experiments.

## 9  Conclusion

We propose an automated technique to generate polynomial quantitative invariants for probabilistic programs by Lagrange interpolation. Fixing the degree of loop invariants, our technique can infer polynomial quantitative loop invariants for simple loops. By choosing sampling points carefully, constraints are simplified so that coefficients of loop invariants can be determined. We also develop a counterexample-guided refining heuristics to find coefficients of quantitative loop invariants. We report applications in several case studies.

Our technique does not yet support parameters such as probability in probabilistic choice commands. Such parameters would induce non-linear constraints over coefficients and parameters. SMT solvers however could not find candidate coefficients and parameters as easily. Also, non-determinism is not implemented in our prototype. We plan to address both issues in our future work.

**Acknowledgements.** This work was supported by the Ministry of Science and Technology of Taiwan (103-2221-E-001 -019 -MY3, 103-2221-E-001 -020 -MY3) and the Natural Science Foundation of China (NSFC) under grant No. 61472473, 61428208, 61361136002, the CAS/SAFEA International Partnership Program for Creative Research Teams.